\documentclass[conference]{IEEEtran}
\IEEEoverridecommandlockouts
\usepackage{cite}
\usepackage{amsmath,amssymb,amsfonts}
\usepackage{algorithmic}
\usepackage{graphicx}
\usepackage{textcomp}
\usepackage{xcolor}
\def\BibTeX{{\rm B\kern-.05em{\sc i\kern-.025em b}\kern-.08em
    T\kern-.1667em\lower.7ex\hbox{E}\kern-.125emX}}

\usepackage[acronyms,nonumberlist,nopostdot,nomain,nogroupskip]{glossaries}
\newacronym{quic}{QUIC}{Quick UDP Internet Connections}
\newacronym{3gpp}{3GPP}{3rd Generation Partnership Project}
\newacronym{adc}{ADC}{Analog to Digital Converter}
\newacronym{5g}{5G}{5th Generation}
\newacronym{aimd}{AIMD}{Additive Increase Multiplicative Decrease}
\newacronym{am}{AM}{Acknowledged Mode}
\newacronym{amc}{AMC}{Adaptive Modulation and Coding}
\newacronym{aqm}{AQM}{Active Queue Management}
\newacronym{awgn}{AGWN}{Additive White Gaussian Noise}
\newacronym{balia}{BALIA}{Balanced Link Adaptation}
\newacronym{bdp}{BDP}{Bandwidth-Delay Product}
\newacronym{bf}{BF}{Beamforming}
\newacronym{cc}{CC}{Congestion Control}
\newacronym{pdf}{PDF}{Probability Density Function}
\newacronym{cdf}{CDF}{Cumulative Distribution Function}
\newacronym{icdf}{Inverse CDF}{Inverse Cumulative Distribution Function}
\newacronym{c-v2x}{C-V2X}{Cellular Vehicle-To-Everything}
\newacronym{ci}{CI}{Close-in free space reference}
\newacronym{cn}{CN}{Core Network}
\newacronym{cqi}{CQI}{Channel Quality Information}
\newacronym{cp}{CP}{Control Plane}
\newacronym{csirs}{CSI-RS}{Channel State Information - Reference Signal}
\newacronym{d2d}{D2D}{Device-to-Device}
\newacronym{dc}{DC}{Dual Connectivity}
\newacronym{dce}{DCE}{Direct Code Execution}
\newacronym{dci}{DCI}{Downlink Control Information}
\newacronym{dl}{DL}{Downlink}
\newacronym{dmr}{DMR}{Deadline Miss Ratio}
\newacronym{dmrs}{DMRS}{DeModulation Reference Signal}
\newacronym{dsrc}{DSRC}{Dedicated Short-Range Communication}
\newacronym{e2e}{E2E}{End-to-End}
\newacronym{ecn}{ECN}{Explicit Congestion Notification}
\newacronym{edf}{EDF}{Earliest Deadline First}
\newacronym{enb}{eNB}{evolved Node Base}
\newacronym{epc}{EPC}{Evolved Packet Core}
\newacronym{es}{ES}{Edge Server}
\newacronym{fdma}{FDMA}{Frequency Division Multiple Access}
\newacronym{fdd}{FDD}{Frequency Division Duplexing}
\newacronym[firstplural=Radio Access Technologies (RATs)]{rat}{RAT}{Radio Access Technology}
\newacronym{fs}{FS}{Fast Switching}
\newacronym{ftp}{FTP}{File Transfer Protocol}
\newacronym{gnb}{gNB}{Next Generation Node Base}
\newacronym{harq}{HARQ}{Hybrid Automatic Repeat reQuest}
\newacronym{hetnet}{HetNet}{Heterogeneous Network}
\newacronym{hh}{HH}{Hard Handover}
\newacronym{hol}{HOL}{Head-of-Line}
\newacronym{ia}{IA}{Initial Access}
\newacronym{imt}{IMT}{International Mobile Telecommunication}
\newacronym{iot}{IoT}{Internet of Things}
\newacronym{lidar}{LiDAR}{Light Detection and Ranging}
\newacronym{los}{LOS}{Line of Sight}
\newacronym{lte}{LTE}{Long Term Evolution}
\newacronym{m2m}{M2M}{Machine to Machine}
\newacronym{mac}{MAC}{Medium Access Control}
\newacronym{mc}{MC}{Multi-Connectivity}
\newacronym{mcs}{MCS}{Modulation and Coding Scheme}
\newacronym{mec}{MEC}{Mobile Edge Cloud}
\newacronym{mi}{MI}{Mutual Information}
\newacronym{mimo}{MIMO}{Multiple Input, Multiple Output}
\newacronym{mmwave}{mmWave}{millimeter wave}
\newacronym{ml}{ML}{machine learning}
\newacronym{mr}{MR}{Maximum Rate}
\newacronym{mss}{MSS}{Maximum Segment Size}
\newacronym{mtd}{MTD}{Machine-Type Device}
\newacronym{mtu}{MTU}{Maximum Transmission Unit}
\newacronym{nn}{NN}{Neural Network}
\newacronym{nsf}{NSF}{National Science Foundation}
\newacronym{nfv}{NFV}{Network Function Virtualization}
\newacronym{nlos}{NLOS}{Non Line of Sight}
\newacronym{nr}{NR}{New Radio}
\newacronym{ofdm}{OFDM}{Orthogonal Frequency Division Multiplexing}
\newacronym{pc}{PC}{Point Cloud}
\newacronym{pdcch}{PDCCH}{Physical Downlonk Control Channel}
\newacronym{pdcp}{PDCP}{Packet Data Convergence Protocol}
\newacronym{pdsch}{PDSCH}{Physical Downlink Shared Channel}
\newacronym{pdu}{PDU}{Packet Data Unit}
\newacronym{pf}{PF}{Proportional Fair}
\newacronym{pgw}{PGW}{Packet Gateway}
\newacronym{phy}{PHY}{Physical}
\newacronym{pbch}{PBCH}{Physical Broadcast Channel}
\newacronym[plural=\gls{mme}s,firstplural=Mobility Management Entities (MMEs)]{mme}{MME}{Mobility Management Entity}
\newacronym{prb}{PRB}{Physical Resource Block}
\newacronym{pss}{PSS}{Primary Synchronization Signal}
\newacronym{pucch}{PUCCH}{Physical Uplink Control Channel}
\newacronym{pusch}{PUSCH}{Physical Uplink Shared Channel}
\newacronym{qos}{QoS}{Quality of Service}
\newacronym{rach}{RACH}{Random Access Channel}
\newacronym{ran}{RAN}{Radio Access Network}
\newacronym{red}{RED}{Random Early Detection}
\newacronym{rf}{RF}{Radio Frequency}
\newacronym{rlc}{RLC}{Radio Link Control}
\newacronym{rlf}{RLF}{Radio Link Failure}
\newacronym{rrc}{RRC}{Radio Resource Control}
\newacronym{rrm}{RRM}{Radio Resource Management}
\newacronym{rr}{RR}{Round Robin}
\newacronym{rs}{RS}{Remote Server}
\newacronym{rsrp}{RSRP}{Reference Signal Received Power}
\newacronym{rss}{RSS}{Received Signal Strength}
\newacronym{rtt}{RTT}{Round Trip Time}
\newacronym{rw}{RW}{Receive Window}
\newacronym{rx}{RX}{Receiver}
\newacronym{sa}{SA}{standalone}
\newacronym{sack}{SACK}{Selective Acknowledgment}
\newacronym{sap}{SAP}{Service Access Point}
\newacronym{sch}{SCH}{Secondary Cell Handover}
\newacronym{scoot}{SCOOT}{Split Cycle Offset Optimization Technique}
\newacronym{sdma}{SDMA}{Spatial Division Multiple Access}
\newacronym{sinr}{SINR}{Signal to Interference plus Noise Ratio}
\newacronym{sm}{SM}{Saturation Mode}
\newacronym{snr}{SNR}{Signal to Noise Ratio}
\newacronym{son}{SON}{Self-Organizing Network}
\newacronym{ss}{SS}{Synchronization Signal}
\newacronym{srs}{SRS}{Sounding Reference Signal}
\newacronym{sss}{SSS}{Secondary Synchronization Signal}
\newacronym{tb}{TB}{Transport Block}
\newacronym{tcp}{TCP}{Transmission Control Protocol}
\newacronym{udp}{UDP}{User Datagram Protocol}
\newacronym{fr1}{FR1}{Frequency Range 1}
\newacronym{fr2}{FR2}{Frequency Range 2}
\newacronym{tdd}{TDD}{Time Division Duplexing}
\newacronym{tdma}{TDMA}{Time Division Multiple Access}
\newacronym{tfl}{TfL}{Transport for London}
\newacronym{thz}{THz}{Terahertz}
\newacronym{tm}{TM}{Transparent Mode}
\newacronym{trp}{TRP}{Transmitter Receiver Pair}
\newacronym{tti}{TTI}{Transmission Time Interval}
\newacronym{ttt}{TTT}{Time-to-Trigger}
\newacronym{tx}{TX}{Transmitter}
\newacronym{ue}{UE}{User Equipment}
\newacronym{ul}{UL}{Uplink}
\newacronym{uml}{UML}{Unified Modeling Language}
\newacronym{um}{UM}{Unacknowledged Mode}
\newacronym{utc}{UTC}{Urban Traffic Control}
\newacronym{v2v}{V2V}{Vehicle-to-Vehicle}
\newacronym{vm}{VM}{Virtual Machine}
\newacronym{rsrq}{RSRQ}{Reference Signal Received Quality}
\newacronym{rssi}{RSSI}{Received Signal Strength Indicator}
\newacronym{crs}{CRS}{Cell Reference Signal}
\newacronym{comp}{CoMP}{Coordinated Multi-Point}
\newacronym{cran}{C-RAN}{Cloud \acrlong{ran}}
\newacronym{ca}{CA}{Carrier Aggregation}
\newacronym{cco}{CC}{Carrier Component}
\newacronym{nsa}{NSA}{Non Stand Alone}
\newacronym{embb}{eMBB}{Enhanced Mobility Broadband}
\newacronym{bsr}{BSR}{Buffer Status Report}
\newacronym{srb}{SRB}{Service Radio Bearer}
\newacronym{scm}{SCM}{Spatial Channel Model}
\newacronym{sctp}{SCTP}{Stream Control Transmission Protocol}
\newacronym{mptcp}{MPTCP}{Multi-path TCP}
\newacronym{ietf}{IETF}{Internet Engineering Task Force}
\newacronym{os}{OS}{Operating System}
\newacronym{tls}{TLS}{Transport Layer Security}
\newacronym{rfc}{RFC}{Request for Comments}
\newacronym{http}{HTTP}{HyperText Transfer Protocol}
\newacronym{nat}{NAT}{Network Address Translation}
\newacronym{api}{API}{Application Programming Interface}
\newacronym{rto}{RTO}{Retransmission Timeout}
\newacronym{psc}{PSC}{Public Safety Communication}
\newacronym{rpgm}{RPGM}{Reference Point Group Mobility}
\newacronym{ic}{IC}{Incident Command}
\newacronym{rsu}{RSU}{Road Side Unit}
\newacronym{uav}{UAV}{Unmanned Aerial Vehicle}
\newacronym{usa}{U.S.}{United States}
\newacronym{vr}{VR}{Virtual Reality}
\newacronym{iab}{IAB}{Integrated Access and Backhaul}
\newacronym{wlan}{WLAN}{Wireless Local Area Network}
\newacronym{cots}{COTS}{Commercial Off-the-Shelf}
\newacronym{fpga}{FPGA}{Field Programmable Gate Array}
\newacronym{rcn}{RCN}{Research Coordination Network}
\newacronym{abg}{ABG}{Alpha-Beta-Gamma}
\newacronym{fi}{FI}{Floating Intercept}
\newacronym{uas}{UAS}{Unmanned Aerial System}
\newacronym{gps}{GPS}{Global Positioning System}
\newacronym{a2g}{A2G}{air-to-ground}
\newacronym{a2a}{A2A}{air-to-air}
\newacronym{uma}{UMa}{Urban Macro}
\newacronym{umi}{UMi}{Urban Micro}
\newacronym{upa}{UPA}{Uniform Planar Array}
\newacronym{rma}{RMa}{Rural Macro}
\newacronym{inoo}{InOo}{Indoor Open Office}
\newacronym{ple}{PLE}{path loss exponent}
\newacronym{aoa}{AoA}{Angle of Arrival}
\newacronym{aod}{AoD}{Angle of Departure}
\newacronym{toa}{ToA}{Time of Arrival}
\newacronym{mpc}{MPC}{Multi-path Component}
\newacronym{cir}{CIR}{Channel Impulse Response}
\newacronym{rt}{RT}{Ray-tracing}
\newacronym{tc}{TC}{Time Cluster}
\newacronym{sl}{SL}{Spatial Lobe}
\newacronym{svd}{SVD}{Singular Value Decomposition}
\newacronym{6g}{6G}{sixth generation}
\newacronym{ns3}{ns-3}{Network Simulator 3}
\newacronym{fsc}{FS}{Fully Stochastic}
\newacronym{hbc}{HB}{Hybrid}
\newacronym{hpbw}{HPBW}{Half Power Beamwidth}
\newacronym{hsc}{HSC}{Hybrid Semantic Compression}
\newacronym{prr}{PRR}{Packet Receipt Rate}
\newacronym{v2x}{V2X}{Vehicle-To-Everything}

\newacronym{ai}{AI}{artificial intelligence}

\newacronym{pqos}{PQoS}{Predictive Quality of Service}
\newacronym{rl}{RL}{reinforcement learning}
\newacronym{mdp}{MDP}{Markov Decision Process}
\newacronym{fl}{FL}{federated learning}
\newacronym{sgd}{SGD}{Stochastic Gradient Descent}
\newacronym{ddqn}{DDQN}{Double Deep Q-Network}
\newacronym{dql}{DQL}{Double Q-Learning}
\newacronym{td}{TD}{teleoperated driving}
\newacronym{map}{mAP}{mean average precision}
\newacronym{kpi}{KPI}{Key Performance Indicator}
\newacronym{mab}{MAB}{Multi-Armed Bandit}
\newacronym{dsarsa}{DSARSA}{Deep SARSA}
\newacronym{sarsa}{SARSA}{State–Action–Reward–State–Action}
\newacronym{qoe}{QoE}{Quality of Experience}

\usepackage[font=small]{caption}
\usepackage[font=footnotesize]{subcaption}

\usepackage[utf8]{inputenc}
\usepackage{pgfplots}
\DeclareUnicodeCharacter{2212}{−}
\usepgfplotslibrary{groupplots,dateplot}
\usetikzlibrary{patterns,shapes.arrows}
\pgfplotsset{compat=newest}

\usepackage{notation}
\usepackage{multirow}

\usepackage{tabularx}
\usepackage{subcaption}

\usepackage[most]{tcolorbox}
\tcbuselibrary{theorems}

\newtcbtheorem{Definitions}{Summary}%
{colframe=black,colback=white,fonttitle=\bfseries}{}

\usetikzlibrary{positioning}

\begin{document}

\title{Federated Reinforcement Learning to Optimize Teleoperated Driving Networks}

\author{\IEEEauthorblockN{{Filippo Bragato, Marco Giordani, Michele Zorzi}\medskip}
\IEEEauthorblockA{
Department of Information Engineering, University of Padova, Italy. \\
Email: \texttt{\{filippo.bragato,marco.giordani,michele.zorzi\}@dei.unipd.it}
\vspace{-0.5cm}\\
}

\thanks{This work was supported by the European Union under the Italian National Recovery and Resilience Plan (NRRP) of NextGenerationEU, partnership on ``Telecommunications of the Future'' (PE0000001 - program ``RESTART'').}}

\maketitle
\begin{abstract}

  Several \gls{6g} use cases have tight requirements in terms of reliability and latency, in particular \gls{td}. 
  To address those requirements, \gls{pqos}, possibly combined with \gls{rl}, has emerged as a valid approach to dynamically adapt the configuration of the \gls{td} application (e.g., the level of compression of automotive data) to the experienced network conditions. 
  In this work, we explore different classes of \gls{rl} algorithms for \gls{pqos}, namely MAB (stateless), SARSA (stateful on-policy), Q-Learning (stateful off-policy), and DSARSA and DDQN (with \gls{nn} approximation). We trained the agents in a \gls{fl} setup to improve the convergence time and fairness, and to promote privacy and security. The goal is to optimize the trade-off between \gls{qos}, measured in terms of the end-to-end latency, and \gls{qoe}, measured in terms of the quality of the resulting compression operation. 
  We show that 
  Q-Learning uses a small number of learnable parameters, and is the best approach to perform \gls{pqos} in the \gls{td} scenario in terms of average reward, convergence, and computational cost.

\end{abstract}

\glsresetall


\begin{tikzpicture}[remember picture,overlay]
  \node[anchor=north,yshift=-10pt] at (current page.north) {\parbox{\dimexpr\textwidth-\fboxsep-\fboxrule\relax}{
      \centering\footnotesize This paper has been accepted for publication at IEEE Global Communications Conference (GLOBECOM). 2024 ©IEEE.\\
      Please cite it as: F. Bragato, M. Giordani, M. Zorzi, “Federated Reinforcement Learning to Optimize Teleoperated Driving Networks,” IEEE Global Communications Conference (GLOBECOM), Cape Town, South Africa, 2024.\\
      }};
\end{tikzpicture}

\glsresetall

\section{Introduction}
\Gls{td} is positioned to revolutionize the automotive industry by enabling remote operations of vehicles with unprecedented levels of precision and safety.
In particular, \gls{td} systems can replace the human control of the vehicle in case of emergency or malfunctions, e.g., during extreme weather conditions or in hostile environments. 
Compared to a fully autonomous driving system, where the control of the vehicle is implemented via software onboard the vehicle itself, in \gls{td} the vehicle sends perceptions of the environment, acquired by videocamera and \gls{lidar} sensors, to a remote host, called teleoperator. The teleoperator, that can be either human or \gls{ai}, receives data from the vehicle, processes it, and sends commands to the vehicle's actuators for proper control~\cite{Georg2018TeleoperatedDriving}.

For \gls{td} to perform safely, the communication between the vehicle and the teleoperator must be reliable and fast. For example, according to the \gls{3gpp} Release-17 specifications, the end-to-end latency should be less than 50 ms \cite{3gpp.22.186}. 
However, the size of a raw \gls{lidar} frame of 82\,200 points is around  1 MB~\cite{testolina2023selma}; with a sensor rate of 30 fps, the resulting data rate is around 240 Mbps, which may be challenging to handle for most~networks.

To solve this issue, \gls{td} data must be ``transformed'' before transmission. 
One option is to compress perceptions before transmission, to reduce the message size and so the link overload. However, 
the compression time is not negligible, especially on \gls{lidar} frames, and depends on the compression algorithm~\cite{nardo2022point}. 
Moreover, compression may significantly deteriorate the quality of the raw perception, with possible negative implications for the teleoperator's performance.

In this context, the research community has been working on \gls{pqos} as a solution to optimize \gls{td} applications~\cite{Boban2021PQoS}. 
In \gls{pqos}, the \gls{td} system receives advance notifications about possible \gls{qos} degradation, and implements dedicated countermeasures to maintain connectivity with the teleoperator.
In our previous works~\cite{Mason2022RLframeworkforPQoS,drago2022CaseStudyns3} we proposed a new \gls{pqos} framework, referred to as RAN-AI, that can identify the optimal compression level for \gls{lidar} data at the application to satisfy network requirements. We proved that RAN-AI can optimize the trade-off between \gls{qos} (measured as the latency between the vehicle and the teleoperator) and \gls{qoe} (measured as the quality of the resulting LiDAR data after compression).\footnote{While \gls{qoe} is formally defined as the degree of delight or annoyance of the user of an application or service, in the telecom community \gls{qoe} is measured as the quality of the application, as opposed to more conventional \gls{qos} metrics that depend on the network, e.g., to measure the effect that some network optimizations produce on the applications. In this paper, we adopt this second definition, and define \gls{qoe} as the quality of the data after compression.}
Specifically, the RAN-AI uses \gls{rl} and, in particular, \gls{ddqn}, for network prediction and optimization, even though this approach requires to learn a huge number of parameters, and is therefore computationally slow to convergence.

Therefore, in this paper we investigate some other classes of \gls{rl} algorithms for \gls{pqos}. We evaluate the trade-off 
between stateless and stateful and between on-policy and off-policy algorithms (where off-policy algorithms, unlike on-policy, use different policies for training and data collection),
and consider both linear and \gls{nn} approximations of the action-value function. Notably, we compare the following algorithms: (i) \gls{mab}, (ii) \gls{sarsa}, (iii) Q-Learning, (iv) \gls{dsarsa}, and (v) \gls{ddqn}.
Following the approach in \cite{bragato2023towards}, we train the \gls{rl} agents in a \gls{fl} setup, which improves the convergence time and fairness compared to centralized \gls{rl}, and promotes privacy and security.
We show via simulations in ns-3 that stateful off-policy outperforms stateless on-policy algorithms, and \gls{nn} approximation is not always better than linear approximation. Finally, we prove that  Q-Learning is the best approach to jointly optimize \gls{qos} and \gls{qoe} for \gls{pqos} in terms of average reward, convergence, and computational cost.

The rest of the paper is organized as follows. In Sec.~\ref{sec:simulation} we describe our PQoS framework, in Sec.~\ref{sec:rl} we present the \gls{rl} algorithms for \gls{pqos}, in Sec.~\ref{sec:results} we show our simulation results, and finally in Sec.~\ref{sec:conclusions} we draw the conclusions.




\section{PQoS Framework} \label{sec:simulation}
Our PQoS framework is based on~\cite{Mason2022RLframeworkforPQoS}, and is implemented in ns-3
via the dedicated \texttt{ns3-ran-ai} module,\footnote{Source code: {https://github.com/signetlabdei/ns3-ran-ai}.} first presented in~\cite{drago2022CaseStudyns3}.
The framework consists of the following modules.

\paragraph{Scenario} 
The scenario consists of $N$ vehicles, referred to as \glspl{ue}, that communicate with the teleoperator.
The teleoperator is a remote host co-located with a \gls{gnb}, so it also acts as the access point to the core network.

\paragraph{Channel and mobility} 
For the sake of realism, the map of the scenario in which the vehicles move and interact is taken from OpenStreetMap (OSM).
Then, the mobility of vehicles is modeled with Simulation of Urban MObility (SUMO)~\cite{SUMO2012}. 
Finally, the wireless channel between the \glspl{ue} and the \gls{gnb} is modeled with Geometry Efficient propagation Model for V2V communication (GEMV2)~\cite{boban2014geometry}. 
Eventually, the channel traces are parsed in ns-3 to get the received power, so as to obtain the actual \gls{qos} of the~link.

\paragraph{Networking} 
Our \texttt{ns3-ran-ai} module for PQoS is built upon the \texttt{ns3-mmwave} module,\footnote{Source code: {https://github.com/nyuwireless-unipd/ns3-mmwavei}.} which is used to simulate the whole 5G NR protocol stack~\cite{mezzavilla2018end}.
We consider \gls{udp} at the transport layer to reduce the protocol overhead and transmit data faster.

\paragraph{Application} 
We consider the transmission of high-quality high-resolution \gls{lidar} frames, produced at a frame rate $f$, from the \glspl{ue} to the teleoperator.
In this work, \gls{lidar} frames are modeled based on SELMA~\cite{testolina2023selma}, a new open-source multimodal synthetic dataset for autonomous driving.\footnote{SELMA dataset: {https://scanlab.dei.unipd.it/selma-dataset/}.} 

\Gls{lidar} frames can be compressed before transmission via Draco, a lossy compression algorithm developed by Google.
This choice is motivated by several claims. First, unlike other solutions, Draco is lightweight and fast for both compression and decompression~\cite{nardo2022point}, and can be executed in real time at \glspl{ue} with limited computational power and with no dedicated hardware.
Second, it implements up to 31 quantization levels and 11 compression levels, for a total of 341 compression configurations, so it gives the agents many opportunities to learn. For example, a more aggressive compression level can reduce the message size, thus improving latency. On the downside, it increases the (de)compression time, and possibly degrades the quality of the resulting~frame. 


\paragraph{RAN-AI} 
It implements an \gls{rl} agent to identify the optimal compression configuration (i.e., the action) for the LiDAR data (and thus the corresponding size of the packets to send), to optimize the trade-off between \gls{qos} (i.e., the communication delay) and \gls{qoe} (i.e., the quality of compression).
While in our previous works the \gls{rl} agent was trained according to the \gls{ddqn} algorithm,  an extended version of the classical Q-Learning, in this paper we study and implement some other possible \gls{rl} solutions, as described in Sec.~\ref{sec:rl}.

\section{\acrlong{rl} Configurations}  \label{sec:rl}

\gls{rl} is a branch of \gls{ml} that focuses on finding an optimal policy $\optimalpolicy$ for an environment to maximize the cumulative future reward~\cite{sutton2018reinforcement}. 
Specifically, the \gls{rl} agent 
works with the assumption that the environment is a \gls{mdp}. As such, the environment is modeled as a tuple $\mathcal{E} = (\S, \A, \P, \R)$, where $\S$ is the set of states, $\A$ is the set of actions, $\P$ is the transition probability matrix, and $\R$ is the reward function. 
At time $t$, the \gls{rl} agent interacts with the environment observing the current state $S_t$, takes action $A_t$, and receives from the environment a reward $R_t$. 
Then, based on $\P$, the environment moves to the next state $S_{t+1}$ where the agent will take the next action $A_{t+1}$.

The goal of the agent is to find the optimal policy $\optimalpolicy$ that maximizes the expected return. The return $G_t$ is the cumulative future reward from time $t$, defined as
\begin{equation}
  G_t = \sum_{\tau=0}^{\infty} \gamma^\tau R_{t+\tau},
\end{equation}
where $\gamma \in (0,1)$ is the discount factor that represents the importance of the future rewards. 

Since in our setup the probability function $\P$ is unknown, we used model-free algorithms to solve the \gls{mdp}, i.e., the \gls{rl} agent(s) collect and use data to approximate the action-value function  $\qpi(s, a)$. In other words, the agent(s) do not have prior knowledge of how their actions will affect the state transitions and/or the rewards.
Specifically, $\qpi(s, a)$, $s\in\S$, $a\in\A$, is a function $\S \times \A \rightarrow \Re$ that maps every pair $(s,a)$ to the expected reward obtained by taking action $a$ in state $s$ and following policy $\pi$. 
Therefore, the goal of the \gls{rl} algorithms is to approximate $\qpi(s, a)$ well enough to derive the optimal policy $\optimalpolicy$.
Most \gls{rl} algorithms must explore the environment to learn $\optimalpolicy$. To guarantee the exploration of the environment, we use an $\e$-greedy policy: at each step, the agent chooses a random action in $a\in\A$ with probability $\e\in[0,1]$ (exploration), and with probability $1-\e$ the action that maximizes the action-value function (exploitation). The value of $\e$ decreases linearly during~training.


For PQoS, \gls{rl} can be a powerful tool to find the optimal compression configuration for LiDAR data, to minimize the communication delay and satisfy quality constraints. 
Given its ability to adapt to the actual network conditions, an \gls{rl} agent can outperform an ``a priori'' choice of  the compression configuration.
In this work, we compare different \gls{rl} algorithms, namely \gls{mab} (Sec.~\ref{sub:mab}), \gls{sarsa} (Sec.~\ref{sub:sarsa}), \gls{dsarsa} (Sec.~\ref{sub:dsarsa}), and Q-Learning and \gls{ddqn} (Sec.~\ref{sub:qlearning}).
Finally, the \gls{rl} algorithms  have been trained in an \gls{fl} setup~\cite{federated}.
Therefore, vehicles periodically share intermediate learning model updates relative to their action-value approximations to a central entity (e.g., co-located at the gNB with the teleoperator), which aggregates the received data into a global model $Q_G(s,a)$, improving on it iteratively.
The resulting global model of the action-value function is then returned to the federated agents, that can further train it on their local data.
 This setup is used to improve the convergence of the \gls{rl} algorithms, and guarantee fairness inside the network~\cite{bragato2023towards}. 

\subsection{\acrfull{mab}}
\label{sub:mab}

\gls{mab} is a stateless RL algorithm that approximates the action-value function $\qpi(s, a)$ as $Q^i_t(a)$, relative to agent $i\in\{1,\dots,N\}$, based solely on action selections and rewards from previous learning steps, and with no explicit knowledge of the underlying state~\cite{kuleshov2014algorithms}.
In our setup, the number of available actions is finite, so $Q^i_t(a)$ 
can be represented by a lookup table. So, \gls{mab} is simple and lightweight.

\emph{Update rule.}
Since the environment is non-stationary, we use the following custom update rule:
\begin{equation}
  Q^i_{t+1}(a) = Q^i_{t}(a) + \lambda\left(R^{i}_t - Q^i_t(a)\right), 
\end{equation}
where $\lambda$ is the learning rate. 
Considering an \gls{fl} setup, learning updates are aggregated into a global model $Q_G(a)$, which is then returned to the local agents for additional training, i.e., 
\begin{equation}
  Q_G(a) = \frac{\sum_{i=1}^N \mathcal{N}^i Q^i_{t+1}(a)}{\sum_{i=1}^N \mathcal{N}^i}, \quad \forall a \in \A,
\end{equation}
where $\mathcal{N}^i$ is the number of learning steps for agent $i$ since the last federated update.

\emph{Summary}. MAB is simple and lightweight, even though it cannot remember previous states.



\subsection{\acrfull{sarsa}}
\label{sub:sarsa}
\Gls{sarsa} is a stateful on-policy \gls{rl} algorithm that evaluates and improves policy $\pi$ until it converges to the optimal policy $\optimalpolicy$. 
Compared to stateless solutions, stateful algorithms also exploit a representation of the environment (i.e., the state) to approximate the action-value function~$\qpi(s, a)$.
The name of \gls{sarsa} comes from the tuple $\{S,A,R,S',A'\}$: in the current state ($S$), an action ($A$) is taken, and the agent gets a reward ($R$) going to the next state ($S'$), where it takes the next action ($A'$). 
SARSA updates the action-value function for the current state based on either the current best action or another, exploratory, action with some (typically small) probability, so it is an example of on-policy learning. 
During exploration, the agent may take suboptimal actions, especially in complex environments, thereby yielding suboptimal results.
Still, with a proper design of the learning process, SARSA has been proved to converge to the optimal policy~\cite{singh2000convergence}.

Since the state is continuous and multidimensional, the action-value function  $\qpi(s, a)$ must be approximated as $\hat Q(\textbf{s}, a, \textbf{w})$, where $\textbf{s}$ ($\textbf{w}$) is a vector of states (weights). The weights are updated at each learning step to minimize the error between $\hat Q(\textbf{s}, a, \textbf{w})$ and $\qpi(s, a)$.
SARSA operates with a linear approximator, which performs an inner product between the state vector (in homogeneous coordinates) and the weights vector related to a specific action, i.e., $\hat Q(\textbf{s}, a, \textbf{w}) = \textbf{s} \cdot \textbf{w}_a$. Therefore, \gls{sarsa} requires a (non-obvious) linear relationship between the state and the action-value function.

\emph{Update rule.}
Let $\delta^{i}_t$ be the difference between the estimate of the action-value function for agent $i\in\{1,\dots,N\}$ at time $t+1$ and that at time $t$,~i.e.,
\begin{equation}
  \delta^{i}_t = R^{i}_{t} - \bar R^{i} + \hat Q(\textbf{s}^{i}_{t+1}, a^{i}_{t+1}, \textbf{w}^{i}_t) - \hat Q(\textbf{s}^{i}_t, a^{i}_t, \textbf{w}^{i}_t),
  \label{eq:error}
\end{equation}
where $\bar R^{i}$ is the average reward collected from the environment.
The update rule can be defined as
\begin{equation}
  \textbf{w}^{i}_{t+1} = \textbf{w}^{i}_t + \alpha \delta^{i}_t \grad \hat Q(\textbf{s}^{i}_t, a^{i}_t, \textbf{w}^{i}_t),
  \label{eq:w}
\end{equation}
where $\alpha$ is the learning rate, and $\grad \hat Q(\textbf{s}_t, a_t, \textbf{w}_t) = \textbf{s}_t$.
The \gls{fl} setup aggregates the weights into a global model $\textbf{w}_G$, where
\begin{equation}
  \textbf{w}_G = \frac{\sum_{i=1}^N \mathcal{N}^i \textbf{w}^i_{t+1}}{\sum_{i=1}^N \mathcal{N}^i},
  \label{eq:update-fl-sarsa}
\end{equation}
that is then returned to the agents for additional local training. 

\emph{Summary.} SARSA is lightweight and guarantees convergence, even though it has limited exploration capabilities.


\subsection{\gls{dsarsa}}
\label{sub:dsarsa}
\gls{dsarsa} builds upon \gls{sarsa}, so it is another example of a stateful on-policy \gls{rl} algorithm. However, \gls{dsarsa} considers an \gls{nn} approximator for  $\qpi(s, a)$, so it can also represent non-linear functions, even though it is computationally more expensive.
In particular,  \gls{dsarsa} uses the state vector $\textbf{s}$ as the input of an \gls{nn} with $\size{\A}$ output neurons (where $\size{\A}$ is the cardinality of the actions space), each of which represents the action-value function of a specific action.
The \gls{nn} has several blocks, each of which consists of a fully connected layer, a ReLU activation function, and a batch normalization~layer. 

\emph{Update rule.}
The \gls{nn} uses the \gls{sgd} algorithm to minimize the error in Eq.~\eqref{eq:error}, and the Adam optimizer to facilitate the convergence of the algorithm. 
The update rule for the \gls{fl} aggregation is as in Eq.~\eqref{eq:update-fl-sarsa}.

\emph{Summary.} \gls{dsarsa} allows non-linear approximation of $\qpi(s, a)$, but is computationally expensive.

\subsection{Q-Learning / \acrfull{ddqn}}
\label{sub:qlearning}

Q-Learning is the stateful off-policy counterpart of SARSA. 
Compared to an on-policy approach, off-policy algorithms learn and improve a policy $\pi'$ while exploring using a different policy.
Even though there is no guarantee on the actual convergence to $\optimalpolicy$, they have been shown to converge faster, in practice, than their on-policy counterparts~\cite{melo2001convergence}.

The main difference between \gls{sarsa} and Q-Learning is in the computation of the error in Eq.~\eqref{eq:error}. 
In SARSA the error is computed using the action performed by the agent at time $t+1$, while in Q-Learning the error is computed using the action that maximizes the action-value function at time $t+1$, regardless of what actions are selected from exploration.
Therefore, for Q-Learning, the error $\delta^i_t$ for agent $i$ at time $t$ is defined as
\begin{equation}
  \delta^i_t = R^i_{t} - \bar R^i + \max_{a \in \A}\Big( \hat Q(\textbf{s}^i_{t+1}, a, \textbf{w}^i_t)\Big) - \hat Q(\textbf{s}^i_t, a^i_t, \textbf{w}^i_t).
  \label{eq:error-q}
\end{equation}

As usual, the action-value function can be approximated by either a linear or an \gls{nn} approximator. In the latter case, the model is trained according to the \gls{ddqn} algorithm described in~\cite{van2016deep}, which is an extended version of the classical Q-Learning approach with improved performance.


\emph{Update rule.} 
The update rule is the same as in Eq.~\eqref{eq:w}, but with the error defined in Eq.~\eqref{eq:error-q}.
The update rule for the \gls{fl} aggregation is as in Eq.~\eqref{eq:update-fl-sarsa}.


\emph{Summary}. Q-Learning and \gls{ddqn} have good and fast exploration capabilities, even though convergence is not guaranteed.

\section{Performance Evaluation}  \label{sec:results}

In Sec.~\ref{sub:simulation_parameters} we describe our simulation parameters, while in Sec.~\ref{sub:experimental_results} we present our experimental results.

\subsection{Simulation Parameters} 
\label{sub:simulation_parameters}

\paragraph{Communication} 
We consider 5G NR communication at 28 GHz, using numerology 3. 
The available bandwidth is 50 MHz, and the transmission power is 28 dBm.

\paragraph{Application} 
We consider LiDAR frames generated at $f=30$ fps. For Draco compression, we consider compression levels $c\in\{0,5,10\}$ using $q\in\{8,9,10\}$ compression bits. Overall, we have 9 compression configurations, represented by a pair $c-q$.
The compression time has been empirically evaluated using the SELMA dataset and an AMD Ryzen Threadripper PRO 3945WX 12-core CPU processor.
The quality of the compression is measured in terms of \gls{map} obtained by the PointPillars object detector~\cite{lang2019pointpillars} on the SELMA dataset after Draco compression. Compression results are reported in Tab.~\ref{tab:draco_stats}.

We deploy up to $N=10$ vehicles. Based on early simulation results, this value is a good compromise to simulate a congested resource-limited scenario, while satisfying the scalability constraints of the ns-3 simulator.

\begin{table}[tb]
  \centering
  \vspace{0.4cm}
  \caption{Compression time and quality using Draco.}
  \label{tab:draco_stats}
  \renewcommand{\arraystretch}{1}
  \begin{tabular}{|c|c|c|c|c|}
  \hline
     $c$ & $q$ & Configuration &Compression time (ms) & \acrshort{map} \\
     \hline
     \multirow{3}{*}{0}     & 8  & 8-00  & 5.17  & 0.257 \\
                            & 9  & 9-00  & 5.22  & 0.580 \\
                            & 10 & 10-00 & 5.50  & 0.686 \\\hline
     \multirow{3}{*}{5}     & 8  & 8-05  & 5.34  & 0.257 \\
                            & 9  & 9-05  & 6.97  & 0.572 \\
                            & 10 & 10-05 & 8.52  & 0.683 \\\hline
     \multirow{3}{*}{10}    & 8  & 8-10  & 8.21  & 0.257 \\
                            & 9  & 9-10  & 9.62  & 0.574 \\
                            & 10 & 10-10 & 11.58 & 0.683 \\\hline
  \end{tabular}
  \vspace{-0.5cm}
\end{table}

\paragraph{Learning} 
\label{par:learning_parameters}
For a fair comparison, we used the same learning setup for all the RL algorithms described in Sec.~\ref{sec:rl}.
The state of the environment is represented by a multidimensional vector $S \in \Re^{18}$, and consists of the following data: 
\begin{itemize}
   \item The average \gls{sinr}, the average number of symbols in the modulation and coding scheme, the average number of subcarriers;
   \item The average, the minimum, the maximum, and the standard deviation of the \acrshort{rlc}, \acrshort{pdcp}, and application delays, and the relative \gls{prr}. 
\end{itemize}

The reward function depends on both \gls{qos} and \gls{qoe} metrics.
The  \gls{qos} is a function of the communication delay $d$ between an agent/vehicle and the teleoperator, while the \gls{qoe} depends on the \gls{map} $m_a$ obtained when using compression configuration $a$ (i.e., the action).
Thus, the reward function is
\begin{equation}
   R=
   \begin{cases}
       m_a & \mbox{if }d<d_{\text{KPI}}, \\
       -d & \mbox{otherwise},
   \end{cases}
\end{equation}
where $d_{\text{KPI}}$ is the latency requirement of the \gls{td} application (set to 50 ms in our simulations~\cite{3gpp.22.186}).
Indeed, the RL algorithms iteratively choose the optimal compression configuration to maximize the QoE without violating the QoS~requirement.

The \gls{rl} algorithms have been trained on SELMA for 1000 episodes of 2400 learning steps, with $\e$ starting from 0.1 and decreasing to 0.0001.
For the NN approximator, we consider 4 blocks with 64, 128, 64, and 16 neurons respectively.

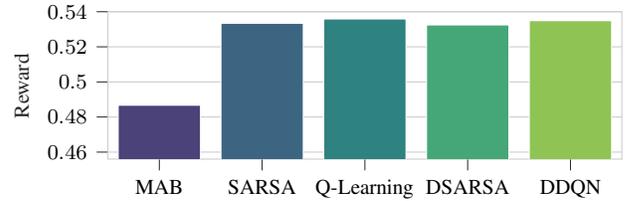
\begin{figure}[tb]
   \centering
   \vspace{0.4cm}
\pgfplotsset{
tick label style={font=\footnotesize},
label style={font=\footnotesize},
legend  style={font=\footnotesize}
}
\begin{tikzpicture}

\definecolor{darkslateblue60101130}{RGB}{60,101,130}
\definecolor{darkslateblue7466121}{RGB}{74,66,121}
\definecolor{darkslategray38}{RGB}{38,38,38}
\definecolor{lightgray204}{RGB}{204,204,204}
\definecolor{mediumseagreen69166119}{RGB}{69,166,119}
\definecolor{seagreen46130127}{RGB}{46,130,127}
\definecolor{yellowgreen14319685}{RGB}{143,196,85}

\begin{axis}[
axis line style={lightgray204},
tick align=outside,
x grid style={lightgray204},
xmajorticks=true,
xtick pos=left,
xmin=-0.5, xmax=4.5,
xtick style={color=darkslategray38},
xtick={0,1,2,3,4},
xticklabels={\gls{mab},SARSA,Q-Learning,\gls{dsarsa},\gls{ddqn}},
y grid style={lightgray204},
ylabel=\textcolor{darkslategray38}{Reward},
ymajorgrids,
ymajorticks=true,
ytick pos=left,
ymin=0.456, ymax=0.54,
ytick style={color=darkslategray38},
height=0.4\linewidth,
width=0.95\linewidth
]
\draw[draw=white,fill=darkslateblue7466121,line width=0.32pt] (axis cs:-0.4,0) rectangle (axis cs:0.4,0.486690586197793); 
\draw[draw=white,fill=darkslateblue60101130,line width=0.32pt] (axis cs:0.6,0) rectangle (axis cs:1.4,0.533388538754397); 
\draw[draw=white,fill=seagreen46130127,line width=0.32pt] (axis cs:1.6,0) rectangle (axis cs:2.4,0.535854593047555); 
\draw[draw=white,fill=mediumseagreen69166119,line width=0.32pt] (axis cs:2.6,0) rectangle (axis cs:3.4,0.532455913344887); 
\draw[draw=white,fill=yellowgreen14319685,line width=0.32pt] (axis cs:3.6,0) rectangle (axis cs:4.4,0.534892882419937); 
\end{axis}

\end{tikzpicture}
   \vspace{-0.9cm} 
   \caption{\vspace{-0.5cm} Average reward at the end of the training. We set $N=5$.}
   \label{fig:reward_average_end}
\end{figure}


\subsection{Experimental Results} 
\label{sub:experimental_results}

\paragraph{Learning results} 
\label{par:learning_results}
We consider $N=5$ vehicles, and compare the different \gls{rl} algorithms in terms of the average reward, which measures the actual performance of the system as it represents the trade-off between \gls{qos} and \gls{qoe}. 
First, in Fig.~\ref{fig:reward_average_end} we plot the reward at the end of the training. We can see that all stateful algorithms outperform stateless \gls{mab}. 
In fact, while MAB makes decisions based solely on the current reward, stateful algorithms maintain an internal state that encodes past experiences, thereby learning patterns, correlations, and dependencies in the environment, and adapting over time.
We also observe that off-policy algorithms (Q-learning and \gls{ddqn}) are slightly better than on-policy algorithms (SARSA and \gls{dsarsa}) since they can better explore the environment from the beginning of the learning phase, even if there is no theoretical guarantee of convergence.

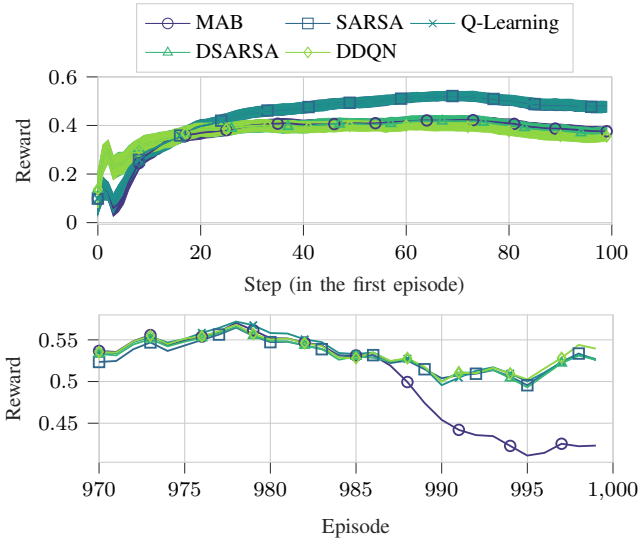
\begin{figure}[t!]
  \centering
  \begin{subfigure}{\linewidth}
    \centering
    \vspace{0.4cm}
    \include{plots/beginning_reward_averaged_errbar}
  \end{subfigure}\vspace{-0.7cm}
  \begin{subfigure}{\linewidth}
    \centering
\pgfplotsset{
tick label style={font=\footnotesize},
label style={font=\footnotesize},
legend  style={font=\footnotesize}
}
\begin{tikzpicture}
\definecolor{QLearningColor}{RGB}{32,144,140}
\definecolor{SARSAColor}{RGB}{48,103,141}
\definecolor{banditColor}{RGB}{68,57,130}
\definecolor{darkslategray38}{RGB}{38,38,38}
\definecolor{lightgray204}{RGB}{204,204,204}
\definecolor{DeepSARSAColor}{RGB}{53,183,120}
\definecolor{DDQLColor}{RGB}{144,214,67}

\begin{axis}[
axis line style={lightgray204},
legend cell align={left},
legend columns=3,
legend style={
  fill opacity=0.8,
  draw opacity=1,
  text opacity=1,
  draw=lightgray204,
  at={(0.5,1.4)},
  anchor=north,
},
tick align=outside,
x grid style={lightgray204},
xlabel=\textcolor{darkslategray38}{Episode},
xmajorgrids,
xmajorticks=true,
xtick pos=left,
xmin=970, xmax=1000,
xtick style={color=darkslategray38},
y grid style={lightgray204},
ylabel=\textcolor{darkslategray38}{Reward},
ymajorgrids,
ymajorticks=true,
ytick pos=left,
ymin=0.403171399439031, ymax=0.579987743136468,
ytick style={color=darkslategray38},
height=0.4\linewidth,
width=0.95\linewidth
]

\addplot [line width=0.7pt, mark=o, mark repeat=3,banditColor, forget plot]
table {%
970 0.536721467971802
971 0.535482883453369
972 0.548418998718262
973 0.555741786956787
974 0.543381929397583
975 0.54930579662323
976 0.553905248641968
977 0.559768438339233
978 0.570341348648071
979 0.56158721446991
980 0.551949381828308
981 0.551781415939331
982 0.546807408332825
983 0.543326139450073
984 0.531276345252991
985 0.531250238418579
986 0.531465768814087
987 0.519229412078857
988 0.499409914016724
989 0.474341034889221
990 0.454081892967224
991 0.442018628120422
992 0.435771226882935
993 0.434537291526794
994 0.422772884368896
995 0.411208510398865
996 0.414469599723816
997 0.425366044044495
998 0.422398209571838
999 0.42317521572113
};
\addplot [line width=0.7pt, mark=square, mark repeat=3,SARSAColor, forget plot]
table {%
970 0.523513555526733
971 0.524652361869812
972 0.538716316223145
973 0.546919465065002
974 0.536699533462524
976 0.549872875213623
977 0.556362867355347
978 0.564717531204224
979 0.554632902145386
980 0.547274351119995
981 0.547768831253052
982 0.543438673019409
983 0.538798332214355
984 0.531540393829346
985 0.529116153717041
986 0.531672954559326
987 0.522184610366821
988 0.525209546089172
989 0.514734029769897
990 0.503990888595581
991 0.507566690444946
992 0.509366512298584
993 0.513700723648071
994 0.506041765213013
995 0.495562553405762
996 0.509955883026123
997 0.523642063140869
998 0.533569574356079
999 0.525668859481812
};
\addplot [line width=0.7pt, mark=x, mark repeat=3,QLearningColor, forget plot]
table {%
970 0.535001039505005
971 0.534098148345947
972 0.547164082527161
973 0.553821444511414
974 0.546792387962341
975 0.551021933555603
976 0.558466911315918
977 0.564326167106628
978 0.571950674057007
979 0.567595243453979
980 0.558214664459229
981 0.557557582855225
982 0.550612330436707
983 0.547261238098145
984 0.534231066703796
985 0.533015251159668
986 0.534852027893066
987 0.524161338806152
988 0.527336239814758
989 0.516211032867432
990 0.495705008506775
991 0.504632711410522
992 0.512902855873108
993 0.517121315002441
994 0.509559631347656
995 0.500105381011963
996 0.510906219482422
997 0.522896528244019
998 0.532747030258179
999 0.526952385902405
};
\addplot [line width=0.7pt, mark=triangle, mark repeat=3,DeepSARSAColor, forget plot]
table {%
970 0.533061265945435
971 0.531391620635986
972 0.544439673423767
973 0.550921440124512
974 0.542265653610229
975 0.547863125801086
976 0.55246901512146
977 0.558209180831909
978 0.565813302993774
979 0.554989099502563
980 0.548641920089722
981 0.548169612884521
982 0.543413162231445
983 0.53908634185791
984 0.526183605194092
985 0.530187606811523
986 0.528491497039795
987 0.524108648300171
988 0.527381062507629
989 0.516944885253906
990 0.500377297401428
991 0.510815382003784
992 0.510517358779907
993 0.513925909996033
994 0.504156589508057
995 0.493085503578186
996 0.507723212242126
997 0.522134304046631
998 0.531044721603394
999 0.526230216026306
};
\addplot [line width=0.7pt, mark=diamond,mark repeat=3, DDQLColor, forget plot]
table {%
970 0.535452604293823
971 0.534808158874512
972 0.548059463500977
973 0.554772138595581
974 0.544962763786316
975 0.551375150680542
976 0.554196357727051
977 0.560441374778748
978 0.5680912733078
979 0.557208776473999
980 0.550530314445496
981 0.551705598831177
982 0.546024322509766
983 0.542662858963013
984 0.528608560562134
985 0.528180837631226
986 0.534957647323608
987 0.524719476699829
988 0.528064966201782
989 0.517506957054138
990 0.499921083450317
991 0.510671019554138
992 0.509968519210815
993 0.51646876335144
994 0.509712219238281
995 0.502570390701294
996 0.515739917755127
997 0.528383016586304
998 0.543997287750244
999 0.539451122283936
};
\end{axis}

\end{tikzpicture}
  \end{subfigure}
  \vspace{-0.8cm} 
  \caption{Average reward over the first 100 steps of the first episode (top) and at the end of the training (bottom), for $N=5$.\vspace{-0.4cm}}
 \label{fig:reward_5}
\end{figure}

To confirm this, in Fig.~\ref{fig:reward_5} (top) we plot the average reward of the \gls{rl} algorithms obtained in the first 100 learning steps of the first episode, that is at the beginning of the training.
We can see that SARSA and Q-Learning are faster to converge than \gls{dsarsa} and \gls{ddqn}. This is because they use a linear approximator for the action-value function, as opposed to a more complex NN, which needs more interactions with the system to start optimizing the network parameters~\cite{pase2022distributed}. 
Interestingly, a \gls{mab} solution, despite its simple design, shows similar performance to \gls{dsarsa} and \gls{ddqn} at the beginning of the training, and is a valid approach to optimize PQoS if the system has little time to learn (e.g., for real-time operations).

Conversely, Fig.~\ref{fig:reward_5} (bottom) shows the evolution of the average reward in the last 30 episodes, that is at the end of the training. As expected, \gls{mab} becomes the worst RL configuration, despite the good training results in the early steps. In fact, as the learning progresses, MAB tends to prioritize exploitation over exploration, and eventually leads to suboptimal outcomes.
In turn, stateful algorithms can continuously adapt and learn from previous states, and iteratively adjust their actions to maximize the long-term reward.
Notably, even though off-policy algorithms are better than on-policy algorithms, the difference is only marginal.
Moreover, the gap between the linear and \gls{nn} approximation is also marginal. This is because the state is not complex, and the reward function is also linear by design, so a linear approximator can accurately represent the action-value function. 
Interestingly, \gls{dsarsa} and \gls{ddqn} eventually converge to the best solution, though after a longer training process.

\begin{table}[b!]
  \centering
  \vspace{-0.3cm}
  \caption{Comparison of the \gls{rl} algorithms. The best is underlined.}
  \label{tab:summary}
  \renewcommand{\arraystretch}{1}
  \begin{tabular}{|r|c|c|c|c|c|}
\hline
     Metric & \gls{mab} & SARSA & Q-learning & \gls{dsarsa} & \gls{ddqn} \\
     \hline
     Reward & 0.486 & 0.533 & \underline{0.535} & 0.532 & \underline{0.535} \\
     Regret & 11.59 & \underline{3.631} & \underline{3.631} & 10.90 & 9.008 \\
     Parameters & \underline{9} & 171 & 171 & 19,529 & 19,529 \\
     Operations & \underline{0} & 333 & 333 & 37,127 & 37,127 \\
     Time & \underline{2.7 $\mu$s} & 0.39 ms & 0.52 ms & 4.4 ms & 5.1 ms \\\hline
  \end{tabular}
\end{table}

In Table~\ref{tab:summary} we summarize the performance of the different RL implementations under several metrics, as follows.
\begin{itemize}
 	\item[(i)] The average reward at the end of the training. As illustrated in Fig.~\ref{fig:reward_average_end}, stateful algorithms have similar performance, even though Q-Learning and \gls{ddqn} eventually outperform the competitors.
 	\item[(ii)] The average regret, defined as the difference between the current reward and the best possible reward obtained in the same number of steps. This metric determines how quickly the \gls{rl} algorithms can learn the optimal policy from a random initialization, so it is an indicator of the learning speed. We can see that SARSA and Q-Learning are fast to optimize (the regret is small), while DSARSA and \gls{ddqn} are slower because of the underlying NN. The performance of MAB is penalized by the simple setup, and the resulting poor reward as the training progresses.
 	\item[(iii)] The learnable parameters. DSARSA and DDQN have to learn many training parameters, which justifies the slower convergence, due to the complexity of the NN. In turn, SARSA and Q-Learning only have to learn the vector of weights for each state and action, and the gap with DSARSA and DDQN is at least two orders of~magnitude.
 	\item[(iv)] The number of arithmetic operations to compute the action-value function in a given state, which is proportional to the computational resources to run the algorithm. For tabular methods like Q-Learning and SARSA, the number of arithmetic operations per update scales linearly with the number of actions, and for MAB it is zero since the action-value function is simply retrieved from a lookup table. For DSARSA and DDQN, instead, the number of operations depends on the complexity of the NN (especially the number of layers and neurons) and on the size of the input space. Notably, DSARSA and DDQN have the same complexity since they use the same NN architecture, and they only differ in the update rule.
 	\item[(v)] The (empirical) average time to perform a learning step. We can see that MAB is the fastest, as it almost does not require any computation to update the action-value function. In turn, SARSA and Q-Learning are faster than DSARSA and DDQN, since the latter two require the computation of the NN forward and backward passes.
 \end{itemize} 

\begin{figure}[t!]
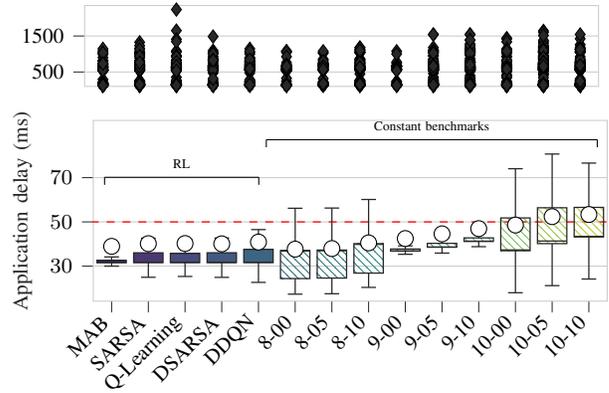

  \centering
  \begin{subfigure}{\linewidth}
    \centering
    \vspace{0.4cm}
    \include{plots/boxplot_outliers_cropped}
  \end{subfigure}\vspace{-0.7cm}
  \begin{subfigure}{\linewidth}
    \centering
   \include{plots/boxplot_time10_users}
  \end{subfigure}
  \vspace{-1.1cm} 
  \caption{Average application delay with $N=10$. Plain (striped) bars are for the RL schemes (constant benchmarks).\vspace{-0.5cm} }
 \label{fig:boxplot_delay_10}
\end{figure}

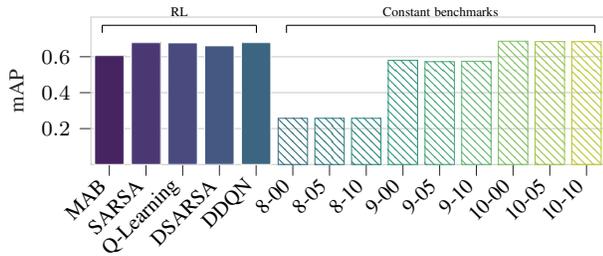
\begin{figure}[t!]
 \centering
\pgfplotsset{
tick label style={font=\footnotesize},
label style={font=\footnotesize},
legend  style={font=\footnotesize}
}
\begin{tikzpicture}

\definecolor{darkseagreen12719397}{RGB}{127,193,97}
\definecolor{darkslateblue60101130}{RGB}{60,101,130}
\definecolor{darkslateblue6789129}{RGB}{67,89,129}
\definecolor{darkslateblue7375126}{RGB}{73,75,126}
\definecolor{darkslateblue7456115}{RGB}{74,56,115}
\definecolor{darkslategray38}{RGB}{38,38,38}
\definecolor{indigo703597}{RGB}{70,35,97}
\definecolor{lightgray204}{RGB}{204,204,204}
\definecolor{mediumseagreen51151124}{RGB}{51,151,124}
\definecolor{mediumseagreen69166119}{RGB}{69,166,119}
\definecolor{mediumseagreen97183112}{RGB}{97,183,112}
\definecolor{seagreen45136126}{RGB}{45,136,126}
\definecolor{seagreen48124128}{RGB}{48,124,128}
\definecolor{seagreen54113129}{RGB}{54,113,129}
\definecolor{yellowgreen15719874}{RGB}{157,198,74}
\definecolor{yellowgreen18920051}{RGB}{189,200,51}

\begin{axis}[
axis line style={lightgray204},
tick align=outside,
x grid style={lightgray204},
xmajorticks=true,
xtick pos=left,
xmin=-0.5, xmax=13.5,
xtick style={color=darkslategray38},
xtick={0,1,2,3,4,5,6,7,8,9,10,11,12,13},
xticklabel style={rotate=45.0,anchor=east},
xticklabels={\gls{mab},SARSA,Q-Learning,\gls{dsarsa},\gls{ddqn},8-00,8-05,8-10,9-00,9-05,9-10,10-00,10-05,10-10},
y grid style={lightgray204},
ylabel=\textcolor{darkslategray38}{\gls{map}},
ytick={0.2,0.4,0.6},
ymajorgrids,
ymajorticks=true,
ytick pos=left,
ymin=0, ymax=0.82071685,
ytick style={color=darkslategray38},
height=0.4\linewidth,
width=0.95\linewidth
]
\draw[draw=white,fill=indigo703597,line width=0.32pt] (axis cs:-0.4,0) rectangle (axis cs:0.4,0.606702618239534);
\draw[draw=white,fill=darkslateblue7456115,line width=0.32pt] (axis cs:0.6,0) rectangle (axis cs:1.4,0.678471396303825);
\draw[draw=white,fill=darkslateblue7375126,line width=0.32pt] (axis cs:1.6,0) rectangle (axis cs:2.4,0.677497761919576);
\draw[draw=white,fill=darkslateblue6789129,line width=0.32pt] (axis cs:2.6,0) rectangle (axis cs:3.4,0.661081358866012);
\draw[draw=white,fill=darkslateblue60101130,line width=0.32pt] (axis cs:3.6,0) rectangle (axis cs:4.4,0.678667824853191);
\draw[draw=seagreen54113129,pattern=north west lines,pattern color=seagreen54113129,line width=0.32pt] (axis cs:4.6,0) rectangle (axis cs:5.4,0.257173);
\draw[draw=seagreen48124128,pattern=north west lines,pattern color=seagreen48124128,line width=0.32pt] (axis cs:5.6,0) rectangle (axis cs:6.4,0.257225);
\draw[draw=seagreen45136126,pattern=north west lines,pattern color=seagreen45136126,line width=0.32pt] (axis cs:6.6,0) rectangle (axis cs:7.4,0.257283);
\draw[draw=mediumseagreen51151124,pattern=north west lines,pattern color=mediumseagreen51151124,line width=0.32pt] (axis cs:7.6,0) rectangle (axis cs:8.4,0.580288);
\draw[draw=mediumseagreen69166119,pattern=north west lines,pattern color=mediumseagreen69166119,line width=0.32pt] (axis cs:8.6,0) rectangle (axis cs:9.4,0.572479000000001);
\draw[draw=mediumseagreen97183112,pattern=north west lines,pattern color=mediumseagreen97183112,line width=0.32pt] (axis cs:9.6,0) rectangle (axis cs:10.4,0.574983000000001);
\draw[draw=darkseagreen12719397,pattern=north west lines,pattern color=darkseagreen12719397,line width=0.32pt] (axis cs:10.6,0) rectangle (axis cs:11.4,0.686397);
\draw[draw=yellowgreen15719874,pattern=north west lines,pattern color=yellowgreen15719874,line width=0.32pt] (axis cs:11.6,0) rectangle (axis cs:12.4,0.683906);
\draw[draw=yellowgreen18920051,pattern=north west lines,pattern color=yellowgreen18920051,line width=0.32pt] (axis cs:12.6,0) rectangle (axis cs:13.4,0.683945000000001);
\end{axis}
 \draw (0.15,1.85) -- node[below] {} node[above] {\tiny RL} ++(2.05,0);
 \draw (0.15,1.85) -- node[below] {} node[above] {} ++(0,-0.1);
 \draw (2.2,1.85) -- node[below] {} node[above] {} ++(0,-0.1);

  \draw (2.6,1.85) -- node[below] {} node[above] {\tiny Constant benchmarks} ++(4.1,0);
 \draw (2.6,1.85) -- node[below] {} node[above] {} ++(0,-0.1);
 \draw (6.7,1.85) -- node[below] {} node[above] {} ++(0,-0.1);
\end{tikzpicture}
 \vspace{-1.1cm} 
 \caption{Average mAP with $N=10$. Plain (striped) bars are for the RL schemes (constant benchmarks).\vspace{-0.5cm}}
 \label{fig:boxplot_qoe_10}
\end{figure}

\paragraph{PQoS results}
In this paragraph we analyze the performance of RL for PQoS in terms of the application delay (Fig.~\ref{fig:boxplot_delay_10}) and the mAP (Fig.~\ref{fig:boxplot_qoe_10}) for $N=10$ vehicles. For comparison, we consider constant benchmarks (see Table~\ref{tab:draco_stats}) in which the compression configuration is set a priori and does not change during the simulation.

\begin{figure}[t!]
 \centering
\pgfplotsset{
tick label style={font=\footnotesize},
label style={font=\footnotesize},
legend  style={font=\footnotesize}
}
\begin{tikzpicture}

\definecolor{darkseagreen12719397}{RGB}{127,193,97}
\definecolor{darkslateblue60101130}{RGB}{60,101,130}
\definecolor{darkslateblue6789129}{RGB}{67,89,129}
\definecolor{darkslateblue7375126}{RGB}{73,75,126}
\definecolor{darkslateblue7456115}{RGB}{74,56,115}
\definecolor{darkslategray38}{RGB}{38,38,38}
\definecolor{indigo703597}{RGB}{70,35,97}
\definecolor{lightgray204}{RGB}{204,204,204}
\definecolor{mediumseagreen51151124}{RGB}{51,151,124}
\definecolor{mediumseagreen69166119}{RGB}{69,166,119}
\definecolor{mediumseagreen97183112}{RGB}{97,183,112}
\definecolor{seagreen45136126}{RGB}{45,136,126}
\definecolor{seagreen48124128}{RGB}{48,124,128}
\definecolor{seagreen54113129}{RGB}{54,113,129}
\definecolor{yellowgreen15719874}{RGB}{157,198,74}
\definecolor{yellowgreen18920051}{RGB}{189,200,51}

\begin{axis}[
axis line style={lightgray204},
tick align=outside,
x grid style={lightgray204},
xmajorticks=true,
xtick pos=left,
xmin=-0.5, xmax=13.5,
xtick style={color=darkslategray38},
xtick={0,1,2,3,4,5,6,7,8,9,10,11,12,13},
xticklabel style={rotate=45.0,anchor=east},
xticklabels={\gls{mab},SARSA,Q-Learning,\gls{dsarsa},\gls{ddqn},8-00,8-05,8-10,9-00,9-05,9-10,10-00,10-05,10-10},
y grid style={lightgray204},
ylabel=\textcolor{darkslategray38}{Reward},
ymajorgrids,
ymajorticks=true,
ytick pos=left,
ymin=0, ymax=0.559340975191527,
ytick style={color=darkslategray38},
height=0.4\linewidth,
width=0.95\linewidth
]
\draw[draw=white,fill=indigo703597,line width=0.32pt] (axis cs:-0.4,0) rectangle (axis cs:0.4,0.394603165133323);
\draw[draw=white,fill=darkslateblue7456115,line width=0.32pt] (axis cs:0.6,0) rectangle (axis cs:1.4,0.437467595420502);
\draw[draw=white,fill=darkslateblue7375126,line width=0.32pt] (axis cs:1.6,0) rectangle (axis cs:2.4,0.437087753758555);
\draw[draw=white,fill=darkslateblue6789129,line width=0.32pt] (axis cs:2.6,0) rectangle (axis cs:3.4,0.424972976400712);
\draw[draw=white,fill=darkslateblue60101130,line width=0.32pt] (axis cs:3.6,0) rectangle (axis cs:4.4,0.421713269402996);
\draw[draw=seagreen54113129,pattern=north west lines,pattern color=seagreen54113129,line width=0.32pt] (axis cs:4.6,0) rectangle (axis cs:5.4,0.0997868043727593);
\draw[draw=seagreen48124128,pattern=north west lines,pattern color=seagreen48124128,line width=0.32pt] (axis cs:5.6,0) rectangle (axis cs:6.4,0.0983790248407565);
\draw[draw=seagreen45136126,pattern=north west lines,pattern color=seagreen45136126,line width=0.32pt] (axis cs:6.6,0) rectangle (axis cs:7.4,0.082935598123477);
\draw[draw=mediumseagreen51151124,pattern=north west lines,pattern color=mediumseagreen51151124,line width=0.32pt] (axis cs:7.6,0) rectangle (axis cs:8.4,0.345591524236798);
\draw[draw=mediumseagreen69166119,pattern=north west lines,pattern color=mediumseagreen69166119,line width=0.32pt] (axis cs:8.6,0) rectangle (axis cs:9.4,0.310802011253137);
\draw[draw=mediumseagreen97183112,pattern=north west lines,pattern color=mediumseagreen97183112,line width=0.32pt] (axis cs:9.6,0) rectangle (axis cs:10.4,0.289259124486508);
\draw[draw=darkseagreen12719397,pattern=north west lines,pattern color=darkseagreen12719397,line width=0.32pt] (axis cs:10.6,0) rectangle (axis cs:11.4,0.283994624607022);
\draw[draw=yellowgreen15719874,pattern=north west lines,pattern color=yellowgreen15719874,line width=0.32pt] (axis cs:11.6,0) rectangle (axis cs:12.4,0.192647487077081);
\draw[draw=yellowgreen18920051,pattern=north west lines,pattern color=yellowgreen18920051,line width=0.32pt] (axis cs:12.6,0) rectangle (axis cs:13.4,0.220585399317015);
\end{axis}
 \draw (0.15,1.85) -- node[below] {} node[above] {\tiny RL} ++(2.05,0);
 \draw (0.15,1.85) -- node[below] {} node[above] {} ++(0,-0.1);
 \draw (2.2,1.85) -- node[below] {} node[above] {} ++(0,-0.1);

  \draw (2.6,1.85) -- node[below] {} node[above] {\tiny Constant benchmarks} ++(4.1,0);
 \draw (2.6,1.85) -- node[below] {} node[above] {} ++(0,-0.1);
 \draw (6.7,1.85) -- node[below] {} node[above] {} ++(0,-0.1);
\end{tikzpicture}
 \vspace{-1.1cm} 
 \caption{Average reward after training. Plain (striped) bars are for the RL schemes (constant benchmarks). We set $N=10$.\vspace{-0.5cm}}
 \label{fig:reward_average_end_10}
\end{figure}
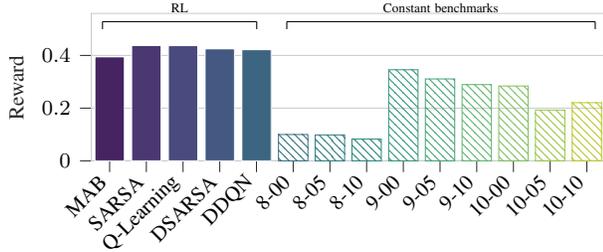

We can see that constant benchmark 8-00 outperforms any other solution in terms of delay since in this configuration data is heavily compressed before transmission, thus reducing the size of the packets to send. However, the resulting mAP is very low. 
In turn, constant benchmark 10-10 maximizes the mAP as it compresses less, but the resulting delay is almost two times higher than that of 8-00.
Notably, RL tries to adapt the compression level to the conditions of the scenario, and achieves the best trade-off between delay (QoS) and mAP (QoE). Most importantly, the application delay is always lower than the TD requirement of 50 ms~~\cite{3gpp.22.186} (represented as a dashed line in Fig.~\ref{fig:boxplot_delay_10}), with no degradation in terms of mAP.
This is confirmed by the reward in Fig.~\ref{fig:reward_average_end_10}, which is slightly lower than for $N=5$ (in Fig.~\ref{fig:reward_average_end}) since the network is more congested. As expected, every RL scheme, including MAB, outperforms the constant benchmarks, meaning that the adaptive behavior of RL is desirable for PQoS.

Even so, RL comes with several outliers (see the upper part of Fig.~\ref{fig:boxplot_delay_10}) where the delay is $\gg$50 ms, which motivates further research towards more advanced solutions that optimize PQoS relative to every packet transmission, and not only on average.

\paragraph{Design guidelines}
SARSA, Q-learning, DSARSA, and DDQN have a similar reward, while MAB does not converge well.
Still, DSARSA and DDQN use an NN approximator for the training, which increases the complexity of the system with limited or no improvements for PQoS. 
The choice lies between SARSA and Q-learning, with a preference on the latter since it provides the best average reward.

\section{Conclusions and Future Work} \label{sec:conclusions}

In this work, we compared different \gls{rl} algorithms to perform \gls{pqos} in a \gls{td} scenario. 
To do so, the agents learn to optimize the compression level of LiDAR data to reduce the transmission delay, without sacrificing the accuracy of the data.
We considered different classes of \gls{rl} algorithms, i.e., \gls{mab}, SARSA, Q-Learning, \gls{dsarsa}, and \gls{ddqn}, trained in an \gls{fl} setup to improve convergence and fairness.

We ran simulations in ns-3 to ensure realistic results, and proved that stateful algorithms outperform \gls{mab}, and off-policy algorithms with linear approximators are slightly better than on-policy algorithms. 
Notably, Q-Learning stands out as a valid, simple, and fast approach to optimize QoS and QoE. 
Still, it suffers from many outliers in the training, so the performance is optimized only on average.

In the future we will implement a multi-layer PQoS approach to jointly optimize the application and the RAN, and design new RL solutions to minimize outliers in the training.

\bibliography{ref}{}
\bibliographystyle{IEEEtran}

\end{document}